\documentclass[%
 aip,
 apl,
 amsmath,amssymb,
 reprint,
]{revtex4-2}

\usepackage{graphicx}
\usepackage{dcolumn}
\usepackage{bm}
\usepackage{xcolor}

\usepackage[utf8]{inputenc}
\usepackage[T1]{fontenc}
\usepackage{mathptmx}
\usepackage{etoolbox}

\usepackage{multirow}

\usepackage[english]{babel}

\makeatletter
\def\@email#1#2{%
 \endgroup
 \patchcmd{\titleblock@produce}
  {\frontmatter@RRAPformat}
  {\frontmatter@RRAPformat{\produce@RRAP{*#1\href{mailto:#2}{#2}}}\frontmatter@RRAPformat}
  {}{}}%
\makeatother

\begin{document}

\preprint{AIP/123-QED}

\title[Draft]{Enhanced electron-beam lithography to reduce the frequency scatter of 200--400\,GHz superconducting microstrip resonators for on-chip filterbank spectrometers}

\author{L.G.G. Olde Scholtenhuis}
\email{L.G.G.OldeScholtenhuis@tudelft.nl}
\affiliation{Delft University of Technology, Electrical Engineering, Mathematics and Computer Science, Delft, the Netherlands}

\author{D.J. Thoen}
\affiliation{SRON — Space Research Organisation Netherlands, Leiden, The Netherlands}

\author{J. J. A. Baselmans}
\affiliation{Delft University of Technology, Electrical Engineering, Mathematics and Computer Science, Delft, the Netherlands}
\affiliation{SRON — Space Research Organisation Netherlands, Leiden, The Netherlands}
\affiliation{Universität zu Köln, Physikalisches Institut, Köln, Germany}

\author{K. Karatsu}
\affiliation{SRON — Space Research Organisation Netherlands, Leiden, The Netherlands}

\author{L.H. Marting}
\affiliation{Delft University of Technology, Electrical Engineering, Mathematics and Computer Science, Delft, the Netherlands}
\affiliation{SRON — Space Research Organisation Netherlands, Leiden, The Netherlands}

\author{S. Vollebregt}
\affiliation{Delft University of Technology, Electrical Engineering, Mathematics and Computer Science, Delft, the Netherlands}

\author{A. Endo}
\affiliation{Delft University of Technology, Electrical Engineering, Mathematics and Computer Science, Delft, the Netherlands}

\date{\today}

\begin{abstract}
Integrated superconducting spectrometers provide the instantaneous bandwidth, sensitivity, and scalable architecture for large-scale spectroscopic surveys in submillimeter-wave astronomy and cosmology.
However, the accuracy with which the resonant frequencies of superconducting microstrip band pass filters can be spaced, has limited the spectral resolution for these spectrometers with continuous spectral coverage to $F/\Delta F < 500$.
The origin of this frequency scatter has been largely unknown.
In this work, we demonstrate a four-fold improvement in the frequency spacing of superconducting microstrip resonators by optimizing electron-beam lithography.
We find that reducing the electron beam step size also reduces the random frequency scatter, and that avoiding main-field stitching across filter patterns can eliminate a systematic frequency shift between groups of resonators, indicating the different origins of these two modes of frequency deviation.
These findings demonstrate that nanometer-scale control of lithographic processes is imperative for the realization of the next-generation integrated superconducting spectrometers with higher spectral resolution.
\end{abstract}

\maketitle

Ultra-wideband spectroscopy at millimeter-submillimeter (mm-submm) wavelengths (100--1000 GHz in frequency), where the instantaneous bandwidth spans an octave or even larger, enables rapid measurement of the redshift, gas content, and physical conditions of dusty star-forming galaxies (DSFGs)\cite{hodgeHighredshiftStarFormation2020} through the detection of molecular and atomic emission lines, tracing galaxy evolution and large-scale structure at high redshift ($z \approx 1$--$10$).
Integrated superconducting spectrometers (ISSs) such as DESHIMA\cite{endoFirstLightDemonstration2019b,karatsuDESHIMA202004002026, moermanAlignmentOpticalVerification2025b, rybakDeshima20Rapid2022},
SuperSpec\cite{karkareFullArrayNoisePerformance2020} and SPT-SLIM\cite{bensonSpectralCharacterizationPerformance2025}
use a superconducting on-chip filterbank and kinetic inductance detectors (KIDs) to provide the required octave-scale bandwidth and photon-noise-limited sensitivity.
The ISS chip is only a few~ $\mathrm{cm^2}$ in size and requires no moving optical components, making it scalable to imaging spectrometers with arrays of spectrometer pixels.
However, unlike quasioptical mm-submm spectrometers (such as grating spectrometers\cite{ferkinhoffZEUS2SecondGeneration2010a, earleZSpecBroadbandDirectdetection2006, liTIMEMillimeterWave2018}, Fourier-transform spectrometers\cite{adeWideFieldofviewLowresolution2020} and Fabry-Pérot interferometers\cite{vavagiakisPrimeCamFirstlightInstrument2018, bradfordSPIFIDirectdetectionImaging2002})
that typically offer guaranteed spectral completeness, filterbank-based ISSs are prone to missing frequency ranges if there are gaps between the passbands of adjacent filters.
Such spectral incompleteness is especially detrimental for single-spatial pixel(spaxel) broadband spectrometers used for redshift measurements (commonly referred to as `Z-
machines'\cite{harrisPurposeBuiltMillimeterWaveRedshift2012a}) 
because each frequency gap translates to a missing redshift range.
Indeed, frequency scatter corresponding to $F/\Delta F \approx 100$--$500$ is common in practice,\cite{martingDirectionalFilterDesign2024a, karatsuDESHIMA202004002026}
making it difficult to match the spectral resolution to the typical emission-line width of massive DSFGs, which corresponds to $F/\Delta F \approx 300$--$3000$.\cite{carilliCOLineWidth2006}


\begin{figure*}[!htbp]
    \includegraphics[width = 0.9\linewidth]{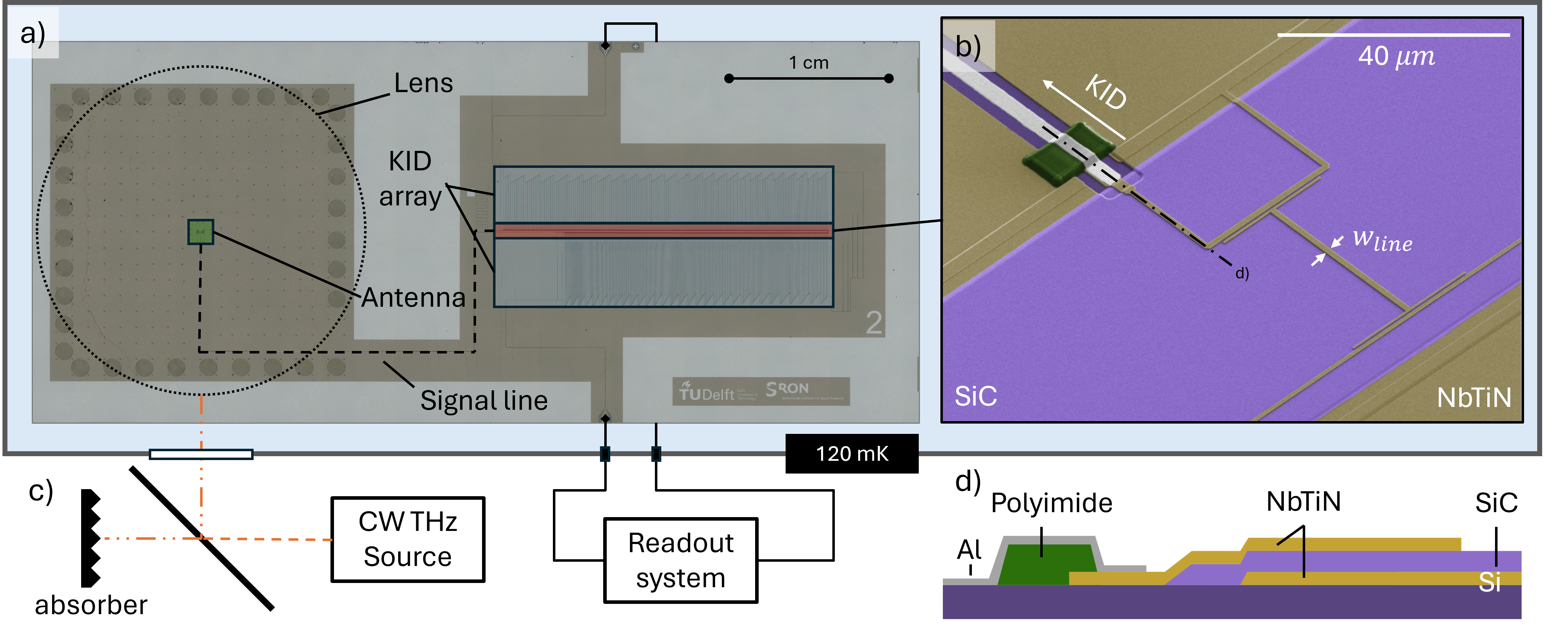}
    \caption{An overview of the device under investigation. a) A microscope image of the chip layout highlighting the various components. The blue area highlights the KID region and the red area contains all the microstrip filters. b) A zoom-in of a single microstrip filter and KID, adapted from Ref.~\onlinecite{pascuallagunaTerahertzBandPassFilters2021a}. The black dashed line indicates the cross-sectional cut shown in d). c) An overview of the measurement setup and its components. The light blue shaded box represents the cryostat.}

    \label{fig:Chip_readout}
\end{figure*}


The minimum frequency spacing between two filters is governed not only by the designed bandwidth but also by the precision of their center transmission frequencies. The precise physical origin of the frequency scatter remains poorly understood, as numerous mechanisms can shift the resonance frequency of a superconducting resonator including variations in layer thickness, kinetic inductance, linewidth, and etching profiles.\cite{martingDirectionalFilterDesign2024a, barryElectromagneticDesignSuperSpec2012, thoenSuperconductingNbTinThin2017a, baselmansKilopixelImagingSystem2017, liReducingFrequencyScatter2022, shuIncreasedMultiplexingSuperconducting2018,mckenneyTileandtrimMicroresonatorArray2019} 
A better understanding of frequency scatter in superconducting filters would also benefit the development of related applications, such as KIDs, given their similarities in materials and fabrication techniques. Alongside efforts to understand these mechanisms, post-fabrication frequency trimming has been proposed for KIDs as a mitigation strategy \cite{liuSuperconductingMicroresonatorArrays2017}, however at mm-submm frequencies the dense spectral packing and small filter dimensions shrink the dimensional tolerances to the alignment accuracy of electron beam patterning ($\approx 10\,\mathrm{nm}$) limiting its applicability. Understanding the origin of frequency scatter is therefore instrumental for on-chip filterbank spectrometers.


In this work, we demonstrate that the lithography-induced frequency scatter in microstrip filters is the dominant factor limiting spectral completeness in our current ISS implementations. We do so by fabricating three 200–400 GHz filterbank spectrometers with systematically varied electron-beam lithography parameters and by quantifying the frequency gaps between adjacent filters. Measuring the electromagnetic response is essential, because scanning electron microscopy (SEM) struggles to detect sub-percent mean line-width variations across lines 800–2000$ \mathrm{nm}$ wide and ~50$\mathrm{\mu m}$ long. This analysis reveals two distinct origins of the frequency gap variations and shows that optimizing these parameters substantially reduces the total frequency scatter.

To investigate the role of electron-beam patterning, we vary two parameters that arise when the mask design is translated into a writable exposure pattern for the electron-beam pattern generator (EBPG). The first is the mainfield layout. Because the EBPG can only deflect the beam over a limited area, large patterns must be divided across multiple mainfields whose boundaries are stitched together by moving the substrate stage. Misalignment at these boundaries can distort feature geometries. The second is the beam step size (BSS), which defines the grid spacing on which the beam places individual exposure points. A coarser BSS limits the fidelity with which the designed geometry is reproduced.

All three devices were fabricated separately but share the same filterbank design. A wideband lens-coupled leaky-wave antenna\cite{dabironezareQuasiOpticalSystemASTE2019a} captures the incoming mm-submm radiation and couples it onto a superconducting microstrip transmission line. Along this line, an array of I-shaped microstrip bandpass filters disperses the radiation by frequency. Each filter is capacitively coupled to a NbTiN/Al hybrid KID\cite{pascuallagunaOnChipSolutionsFuture2022} that senses the transmitted power. The chip layout is shown in Figure~\ref{fig:Chip_readout}a.
The filterbank is designed to cover a total bandwidth of 220--440\,GHz and a spectral resolution of $R = F/\Delta F = 500$, aligned with the scientific goals of the DESHIMA 2.0 project.\cite{karatsuDESHIMA202004002026, endoFirstLightDemonstration2019b, rybakDeshima20Rapid2022}
The filters are arranged on two sides of the central signal line in alternating order: side~A contains filters $F_1, F_3, F_5, \dotsm$ and side~B contains filters $F_2, F_4, F_6, \dotsm$

Each I-shaped filter consists of a center-microstrip, which is coupled to the signal line and KID through two coupler microstrips. The widths of the center and coupler microstrips are identical for all filters and set by impedance-matching to the signal line and KID. Given the thin dielectric and the NbTiN sheet inductance ($1.05\,\mathrm{pH/\square}$), this yields widths of $2000\, \mathrm{nm}$ for the center line and $800\,\mathrm{nm}$ for the coupler lines. The lengths of the center-microstrip and couplers are varied to tune the resonance frequency across the 220–440 GHz band. A more detailed description of the MSL-filter design geometry and parameters is given by Pascual Laguna et al.\cite{pascuallagunaOnChipSolutionsFuture2022,pascuallagunaTerahertzBandPassFilters2021a}

\begin{figure*}[htbp!]
    \includegraphics[width = 0.8\linewidth]{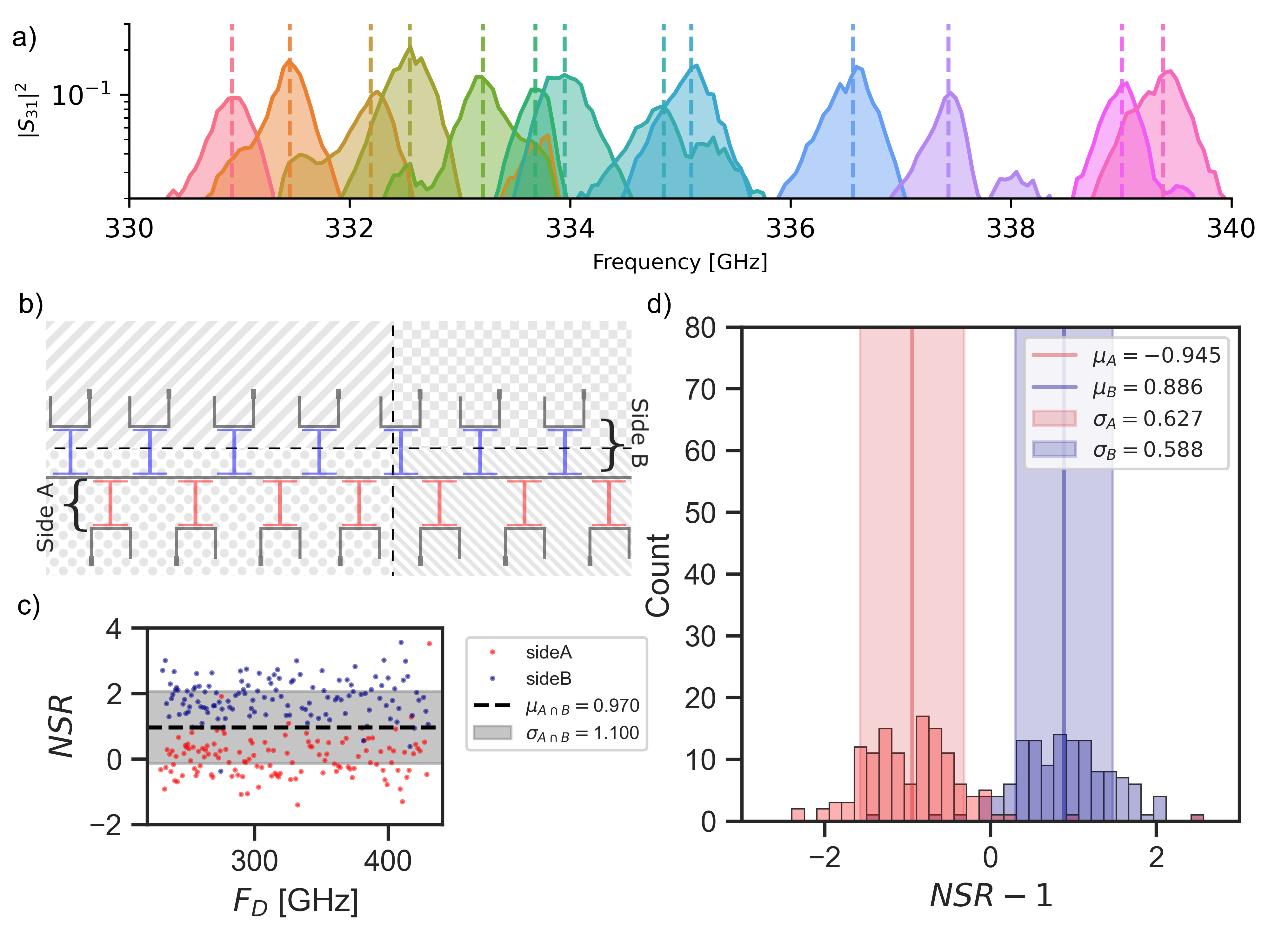}
    \caption{Frequency scatter statistics of D1. a) Filter responses in the frequency range 330–340 GHz. The dashed lines represent the fitted peak frequency of each filter ($F^n_M$). b) A schematic of the writing pattern. The hatched regions indicate different writing mainfields; the black dashed line marks the mainfield boundaries. c) The normalized spacing ratio (NSR) for each inter-filter gap, where NSR = 1 indicates a perfect match to the design. Each point represents the gap between two adjacent filters. The horizontal dashed line shows the mean NSR ($\mu$) and the shaded band its standard deviation ($\sigma$). d) A histogram of NSR - 1, centering the distributions around zero for the case of perfect frequency spacing. Vertical lines show the mean of each side and the shaded areas their standard deviations. In all panels, colors indicate the two sides of the filterbank (side A and side B)}
    \label{fig:F-Scatter_LT223}
\end{figure*}

\begin{figure*}[htbp!]
    \centering
    \includegraphics[width = 0.8\linewidth]{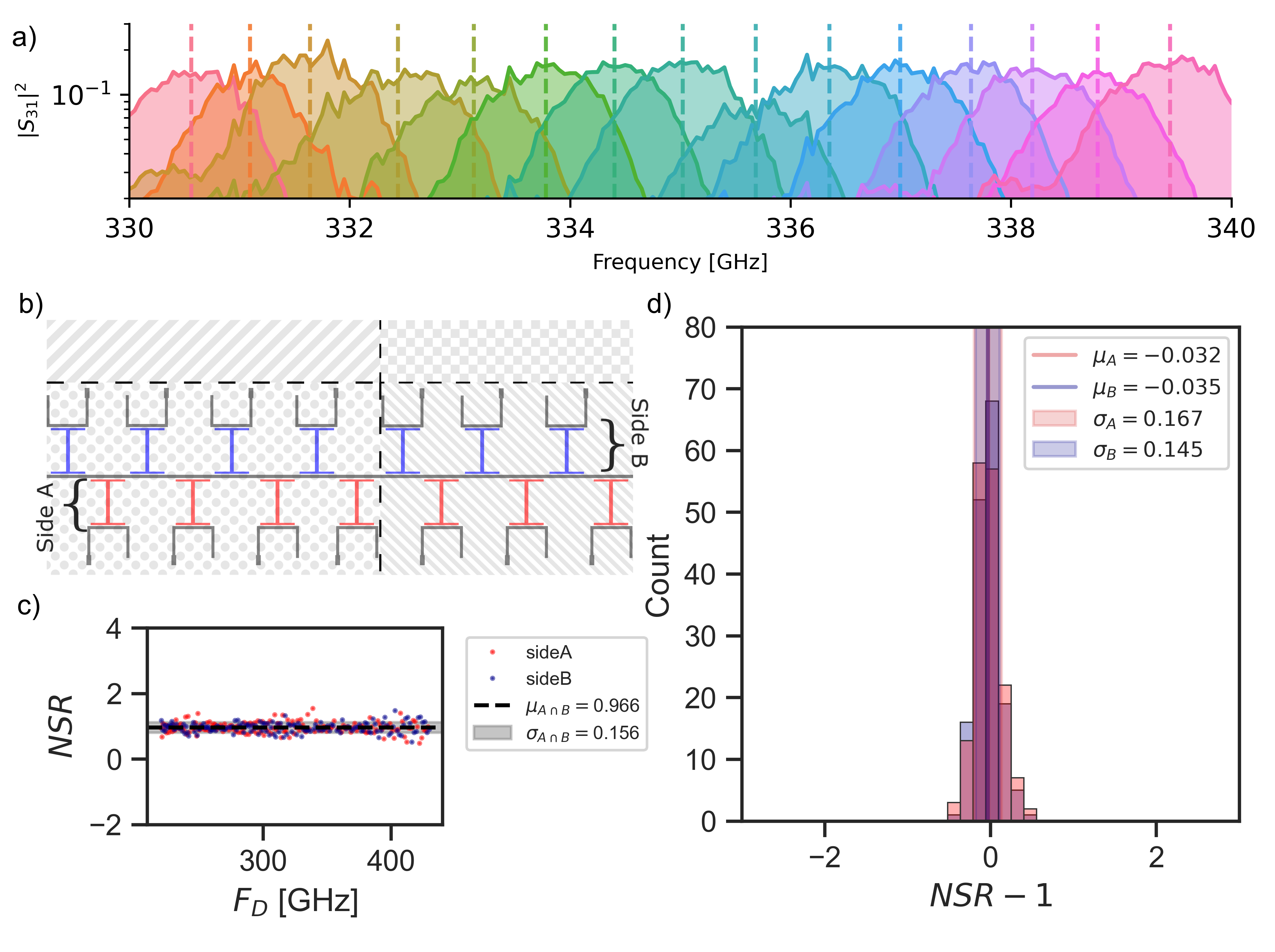}
    \caption{Frequency scatter statistics of D2. a)~Filter responses in the frequency range 330--340\,GHz. The dashed lines represent the fitted peak frequency of each filter ($F^n_M$). b)~A schematic of the writing pattern, showing that all filters are fully contained within single mainfields, eliminating mainfield-edge crossings entirely. c)~The normalized spacing ratio (NSR) for each inter-filter gap, where $\mathrm{NSR} = 1$ indicates a perfect match to the design. Each point represents the gap between two adjacent filters. The horizontal dashed line shows the mean NSR ($\mu$) of the filterbank and the shaded band its standard deviation ($\sigma$). d)~A histogram of NSR$\,-\,1$, centering the distributions around zero for the case of perfect frequency spacing. Vertical lines show the mean of each side and the shaded areas their standard deviations. In all panels, colors indicate the two sides of the filterbank (side~A and side~B).}
    \label{fig:F-Scatter_LT263}
\end{figure*}

We mainly compare Device 1 (D1) and Device 2 (D2), and report the results of Device 3 (D3) in Supplementary Material C, which is a copy of D2 fabricated to confirm the fabrication reproducibility. In D1, as shown if Figure \ref{fig:F-Scatter_LT223}b, one side of the filterbank was patterned such that mainfield boundaries crossed the filter patterns (side B), while each filter on the other side (side A) was fully contained within a single mainfield. In D2, all filters were positioned within single mainfields, eliminating mainfield-edge crossings entirely, see Figure \ref{fig:F-Scatter_LT263}b. The BSS was set to $6.25\,\mathrm{nm}$ for D1 and $1.00\,\mathrm{nm}$ for D2. The writing patterns are illustrated schematically in Figures~\ref{fig:F-Scatter_LT223}b and \ref{fig:F-Scatter_LT263}b. Note that the reduced BSS in D2 increased the writing time by approximately a factor of~40.


All devices have a $200\,\mathrm{nm}$ thick ground layer of NbTiN, deposited using reactive plasma sputtering, similar to the one described by Thoen et al.\cite{thoenSuperconductingNbTinThin2017a}
For D1, this layer was covered with a $500\ \mathrm{nm}$ hydrogenated amorphous silicon (a-Si:H) layer deposited using an Oxford Instruments Plasmapro 80 Plus (PECVD) \cite{buijtendorpLowlossDepositedDielectrics2024} system. D2 and D3, made use of a $500 \ \mathrm{nm}$ amorphous silicon carbide (a-SiC:H) layer. Note that the difference in the relative permittivity between a-Si ($\epsilon_r = 10$) and a-SiC ($\epsilon_r = 7.8$) leads to a \textit{global} offset in the resonance frequency by a factor of $\sqrt{10/7.8}=1.13$, but it has a negligible effect on the \textit{fractional} frequency shift due to lithographic errors studied in this Letter, as confirmed by simulations using as confirmed by simulations using Sonnet \textit{em}. 
The dielectric layer was subsequently covered by another $200\,\mathrm{nm}$ NbTiN layer. \cite{buijtendorpHydrogenatedAmorphousSilicon2022a} For the lithography process, we used the mix-and-match lithography developed by Thoen et al.\cite{thoenCombinedUltravioletElectronbeam2022b} For exposure, we used the Raith 100\,keV EPBG5200 system. The resist was developed in ma-D533s for 45~seconds and dry-etched using a reactive ion etcher (RIE) with an SF$_6$/O$_2$ plasma. End-point detection was used to minimize over-etch of the dielectric layer. This was followed by an in-situ RIE O2 descum, to reduce polyfluor remnants on the resist facilitating resist removal. After this the resist was stripped using 60$^{\circ}$C AZ100 remover.
The hybrid KIDs used in these devices consist of a NbTiN ground plane and a $40\,\mathrm{nm}$ aluminium layer deposited using an Evatec LLS system. The polyimide steps guarantee proper step coverage and contact between the layers.\cite{pascuallagunaOnChipSolutionsFuture2022}
Figure~\ref{fig:Chip_readout}d shows a cross section of the filter and KID.

Following fabrication, we placed each device in a cryostat and measured the KID responses at $120\,\mathrm{mK}$ as function of the frequency of a single mm-wave signal coupled to the lens-antenna of the device via a set of windows and IR filters. The mm-submm signal is generated using a photomixing continuous-wave terahertz source (Toptica Terabeam~1550), schematically shown in Figure~\ref{fig:Chip_readout}c.\cite{endoWidebandOnchipTerahertz2019}. The phase responses of the KIDs were read out using a system analogous to the one described by van Rantwijk et al.\cite{vanrantwijkMultiplexedReadout1000Pixel2016a}.
The individual filter transmission $|S_{31}^i|^2$, shown in Figure \ref{fig:F-Scatter_LT223}a, is obtained from this experiment by dividing the filter KID response ($x_{filter,i}(f)$) with a "wide band KID" response in front of the filterbank ($x_{wb,bf}(f)$) to remove the frequency dependent power fluctuation. We then multiply this with the simulated efficiency of the wide band KID ($\eta_{wb,bf}$), following the method described by Karatsu et al.\cite{karatsuDESHIMA202004002026}, such that $|S_{31}^i|^2 \approx \frac{x_{filter,i}(f)}{x_{wb,bf}(f)} \eta_{wb,bf} $. These were $\eta_{wb,bf} = -30.5\ \mathrm{dB}$ for D1 and $-29\ \mathrm{dB}$ for D2 and D3. Cross-identification of filters with their corresponding KID is described in more detail in Supplementary Material B.

To evaluate the size of the frequency gaps between filters, we introduce the normalized spacing ratio (NSR) defined as  $R(\frac{F_M^{n+1} - F_M^n}{F_M^n})$, where R is the designed spectral resolution of the filterbank, and $F^{i}_M$ the measured peak frequency of the "$i^{th}$" filter. An NSR of 1 represents a measured frequency step that matches the design, while values above or below 1 correspond to frequency steps that are larger or smaller than the design. 

Figure~\ref{fig:F-Scatter_LT223}c shows the NSR for the filters in device D1. A clear separation in the mean NSR is observed between the two sides of the filterbank. This pattern arises because the filter-numbering scheme alternates between side~A and side~B with increasing filter number, so any systematic offset between the two sides appears as a symmetric deviation from the expected mean of $\mu = 1$.

Filters on side~B, whose patterns cross mainfield boundaries, are shifted to lower frequencies relative to those on side~A. This frequency offset is shown more explicitly in Figure~\ref{fig:F-Scatter_LT223}d, where 1 has been subtracted from the NSR to center the distributions around the expected filter separation. Although the means of the two groups differ significantly ($|\mu_A - \mu_B| = 1.831$), their standard deviations are nearly identical ($\sigma_A, \sigma_B \approx 0.6$). This indicates that the mainfield boundary---the only geometrical difference in the fabrication of the two sides---introduces a systematic frequency shift while leaving the random frequency scatter essentially unchanged.


For D2, where the BSS was reduced from $6.25\,\mathrm{nm}$ to $1.00\,\mathrm{nm}$ and all filters were contained within single mainfields, we observe a steep decrease in both the systematic frequency shift and the random frequency scatter (Figure~\ref{fig:F-Scatter_LT263}). The negligible difference in mean between the two sides of the filterbank ($|\mu_A - \mu_B| = 0.003$, see Table~\ref{tab:overview_device_stats}) confirms that mainfield stitching was the root cause of the systematic shift observed in D1. Furthermore, the standard deviation in NSR decreased by approximately $70\%$, from $\sigma \approx 0.53$ (average of sides A and B of D1) to $\sigma \approx 0.16$ for D2. Since the filters in D2 are identical in design to those in D1 except for the BSS and mainfield arrangement, we conclude that the reduced BSS is responsible for the improved frequency scatter.


With D3, fabricated using the same recipes and lithography procedure as D2, we again observed a comparatively small difference in mean between the two sides of the filterbank and a comparable standard deviation. These consistent results suggest a negligible effect of uncontrolled device-to-device fabrication variations on the observed frequency scatter.
The results from D3 are presented in Supplementary Material C. The standard deviations and differences in means of all three devices are summarized in Table~\ref{tab:overview_device_stats}.

\begin{table}[htbp]
\begin{ruledtabular}
\centering
\caption{The averaged measured standard deviation in frequency scatter ($\sigma_i$) of the devices under investigation over multiple measurements. The subscripts A and B indicate the side of the filterbank. The absolute difference of mean frequency between the sides ($|\mu_A - \mu_B|$) is shown in the last column.}
\begin{tabular}{lllll}
Device & $\sigma_{A\cap B}$& $\sigma_A$ & $\sigma_B$ & $ |\mu_A-\mu _B|$ \\
\hline \\
D1 & 1.100          & 0.627    & 0.588    & 1.831  \\
D2 & 0.156          & 0.167    & 0.145    & 0.003  \\
D3 & 0.139          & 0.143    & 0.132    & 0.038      
\end{tabular}
\label{tab:overview_device_stats}
\end{ruledtabular}
\end{table}


The NSR distribution also allows us to estimate the maximum spectral resolution achievable given the observed frequency scatter. After removing the systematic mainfield-stitching offset, the residual NSR values follow an approximately standard normal distribution $\mathcal{N}(0,\,\sigma^2)$. 

We require that the minimum absorbed power ($A_{min}$) for all frequencies should exceed two thirds of the peak absorption frequency of a single filter ($|S_{31}|^2_{max}$) for $95\%$ ($2\sigma$) of the filters. This requirements sets an upper limit on the tolerable standard deviation of $\sigma_\mathrm{max} = 0.4/2 \approx 0.2$. The threshold of $0.4$ corresponds to the situation where the combined signal absorption should at all times exceed $A_{min} = \frac{2}{3} |S_{31}|^2_{max} \approx 0.106$ and is valid when $Q \approx R \gg 1$. The derivation is shown in Supplementary Material A.

Comparing this threshold to the measured values in Table~\ref{tab:overview_device_stats}, both D2 ($\sigma \approx 0.16$) and D3 ($\sigma \approx 0.14$) satisfy the requirement, confirming that the improved lithography process is sufficient to support the design spectral resolution of $R = 500$. By contrast, if we consider only the random scatter of D1 (i.e.\ excluding the systematic mainfield-stitching offset), we obtain $\sigma \approx 0.60$, which corresponds to a scatter-limited spectral resolution of $R = 500\cdot\frac{\sigma_{max}}{0.60}\approx 166$ well below the design target, illustrating the necessity of the lithographic improvements demonstrated here.

To conclude, by fabricating and measuring multiple operating on-chip filterbank spectrometers with different electron-beam lithography parameters, we identified two distinct sources of center-frequency deviation in superconducting microstrip resonators: a systematic shift caused by mainfield stitching, and a random scatter linked to the beam step size. By avoiding mainfield-edge crossings across the filters and reducing the BSS from $6.25\,\mathrm{nm}$ to $1.00\,\mathrm{nm}$, we eliminated the systematic shift and reduced the random frequency scatter by approximately $60\%$. 

The elimination of systematic shifts and the reduction in random scatter together allow for closer packing of filters in the filterbank. Taking both contributions into account, the total observed frequency deviation ($\sigma_{A\cap B}$) is reduced by approximately a factor of 4. Since the spectral resolution scales inversely with frequency deviation, this corresponds to a 4-fold increase in achievable
$R$.

A structured study on the underlying physical mechanisms of the frequency scatter could help further reduce the remaining deviations. For instance, the scatter of Device 2 may no longer be dominated by the electron-beam writing process; other mechanisms such as local variations in kinetic inductance or dielectric constant could now be the limiting factor. Whether the scatter can be further reduced by even finer beam control, or whether a different fabrication parameter now sets the limit, remains an important open question for future work.

Another approach to reducing frequency scatter is to design filter structures that are intrinsically less sensitive to linewidth variations. In the current I-shaped filter, the coupler and center-bar have different linewidths, so a constant linewidth deviation causes a different fractional impedance change on each feature, potentially introducing an impedance mismatch within the filter. A linewidth-uniform geometry, such as the C-shaped filter presented by Pascual Laguna~\cite{pascuallagunaTerahertzBandPassFilters2021a}, offers comparable transmission properties while avoiding this effect, making it a promising direction for future filterbank designs.

These results demonstrate that nanometer-scale control of microstrip-resonator geometries is essential for realizing next-generation integrated superconducting spectrometers with higher spectral density.
As a result of these improvements, compact superconducting on-chip spectrometers are now better positioned to deliver the spectral completeness required for broadband redshift surveys and other astronomical programs that demand gap-free frequency coverage.

\begin{acknowledgments}
We thank the staff of SRON, the Else Kooi Laboratory, and the Kavli Nanolab Delft for their support. This work was supported by the European Union (ERC Consolidator Grant No. 101043486 TIFUUN). Views and opinions expressed are, however, those of the authors only and do not necessarily reflect those of the European Union or the European Research Council Executive Agency. Neither the European Union nor the granting authority can be held responsible for them.
\end{acknowledgments}

\section*{Data Availability Statement}
The data that support the findings of this study are available from the corresponding author upon reasonable request.


\appendix

\section{NSR limit derivation}
\label{app: tolerable sigma}

For a filterbank to provide continuous spectral coverage, the combined absorption between any two adjacent filters must not fall below some minimum threshold. We define a spectral hole as any frequency at which the total absorption drops below $A_\mathrm{min}$, and derive the maximum tolerable filter frequency scatter in terms of this threshold.
For a high spectral resolution filterbank ($R \gg 1$), the absorption minimum between adjacent filters occurs approximately at the midpoint $f_m \approx f_1\left(1 + \frac{1}{2R}\right)$, where $f_1$ is the peak frequency of the lower filter. The individual filter responses at this point are:
$$x_1(f_m) = \frac{|S_{31}^1|^2_{max}}{1+4Q^2\left(\frac{f_{m} - f_1}{f_1}\right)^2} = \frac{|S_{31}^1|^2_{max}}{1+4Q^2\left(\frac{1}{2R}\right)^2}  \equiv x_m$$
$$x_2(f_m) = \frac{|S_{31}^2|^2_{max}}{1+4Q^2\left(\frac{f_{m} - f_2}{f_2}\right)^2_{max}} = \frac{|S_{31}^2|^2_{max}}{1+4Q^2\left(\frac{-1}{2(R+1)}\right)^2} \approx x_m$$
Where we assume equal quality factors $Q$ and peak transmission amplitudes $|S_{31}|^2_{max}$ for adjacent filters, and use $R \gg 1$ so that $\frac{1}{2(R+1)} \approx \frac{1}{2R}$. The combined absorption of the two filters at $f_m$ is then:
$$A(f_{m}) = 1-(1-x_1(f_{m}))(1-x_2(f_{m})) \approx 1 - (1-x_m)^2$$
The natural threshold for continuous spectral coverage is $A_\mathrm{min} = |S_{31}|^2_{max}$, i.e.\ the combined absorption between adjacent filters should nowhere fall below the peak absorption of a single filter. At this threshold, with $Q = R$, the midpoint absorption equals exactly $|S_{31}|^2_{max}$, confirming that $Q = R$ is precisely the design criterion for $-3\,\mathrm{dB}$ filter crossing and corresponds to $\mathrm{NSR} - 1 = 0$. We adopt a slightly relaxed threshold of $A_\mathrm{min} = \frac{2}{3}|S_{31}|^2_{max} \approx 0.106$, which increases the required integration time at the midpoint by at most a factor of $\frac{3}{2}$, giving $\mathrm{NSR} - 1 = 0.4$, as derived below. Given $A_\mathrm{min}$, the corresponding midpoint transmission is:
$$x_m = 1 - \sqrt{1-A_\mathrm{min}}$$
Inverting the filter response equation gives the maximum tolerable fractional frequency offset:
$$\frac{f_m}{f_1} - 1 = \frac{1}{2Q}\sqrt{\frac{|S_{31}|^2_{max}}{x_m}-1}$$
The NSR is defined as the ratio of the actual to designed filter spacing:
$$NSR = R\left(\frac{F_M^{n+1} - F_M^n}{F_M^n}\right)$$
With an expectation value of 1 for a perfectly designed filterbank. To center this around zero we subtract 1. Since $f_m$ lies exactly halfway between $F_M^n = f_1$ and $F_M^{n+1}$, we have $F_M^{n+1} - F_M^n = 2(f_m - f_1)$, giving:
$$NSR - 1 = 2R\left(\frac{f_m}{f_1} - 1\right) - 1$$
For the devices considered here (D1,D2 and D3) with $Q \approx R \approx 500$ and $|S_{31}|^2 \approx 0.16$, the threshold $A_\mathrm{min} = 0.106$ gives $x_m \approx 0.054$, $f_m/f_1 - 1 \approx 0.0014$, and:
$$NSR - 1 = 2(500)(0.0014) - 1 = 0.4$$
This result is used in the main text to define the maximum tolerable filter frequency scatter: requiring that $2\sigma < 0.4$ sets an upper bound of $\sigma_\mathrm{max} \approx 0.2$ on the NSR standard deviation.

\section{Cross identification of KID and filter channel}
\label{app:point_matching}
\begin{figure}[!htbp]
    \centering
    \includegraphics[width=0.9\linewidth]{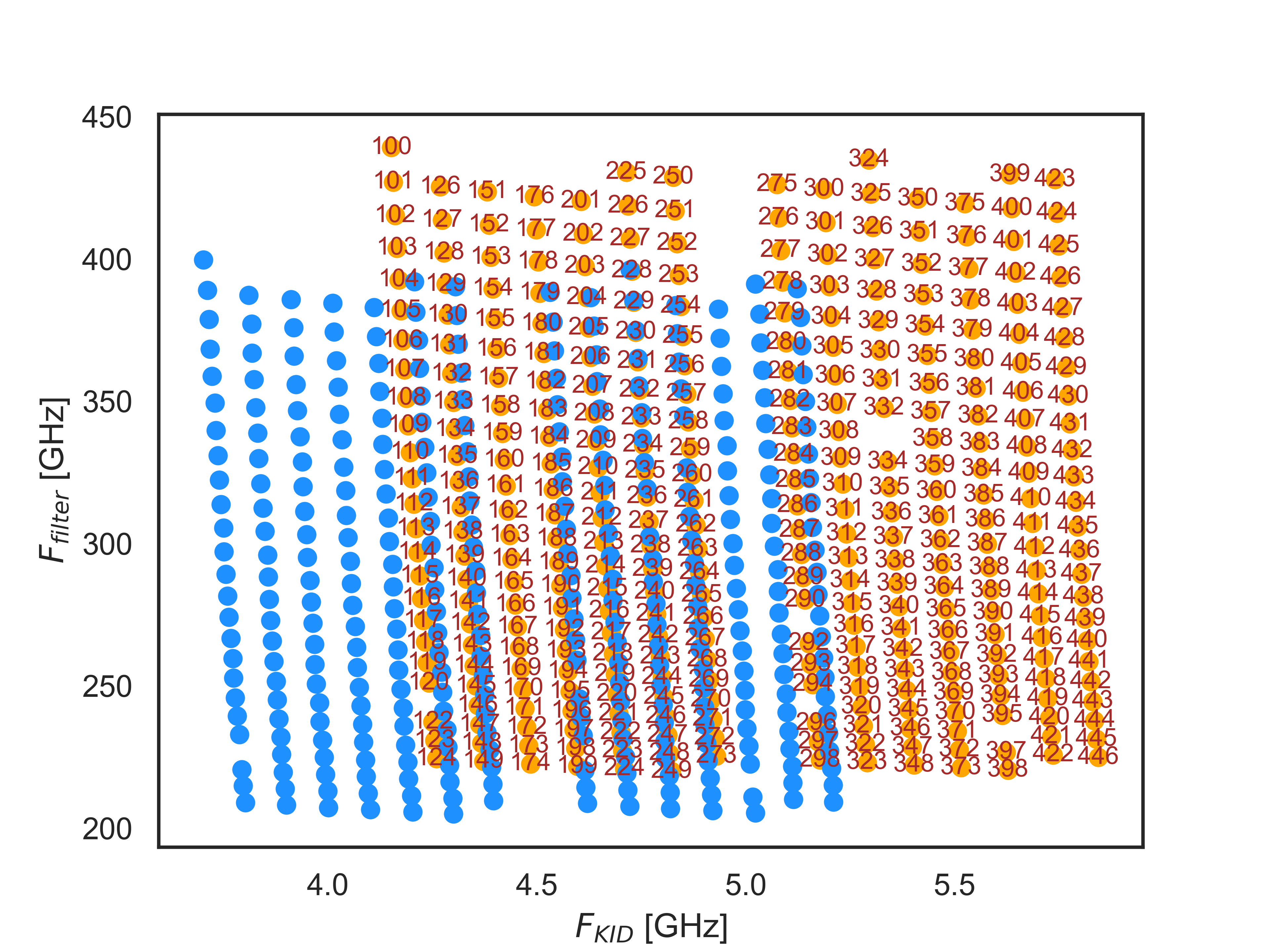}
    \includegraphics[width=0.9\linewidth]{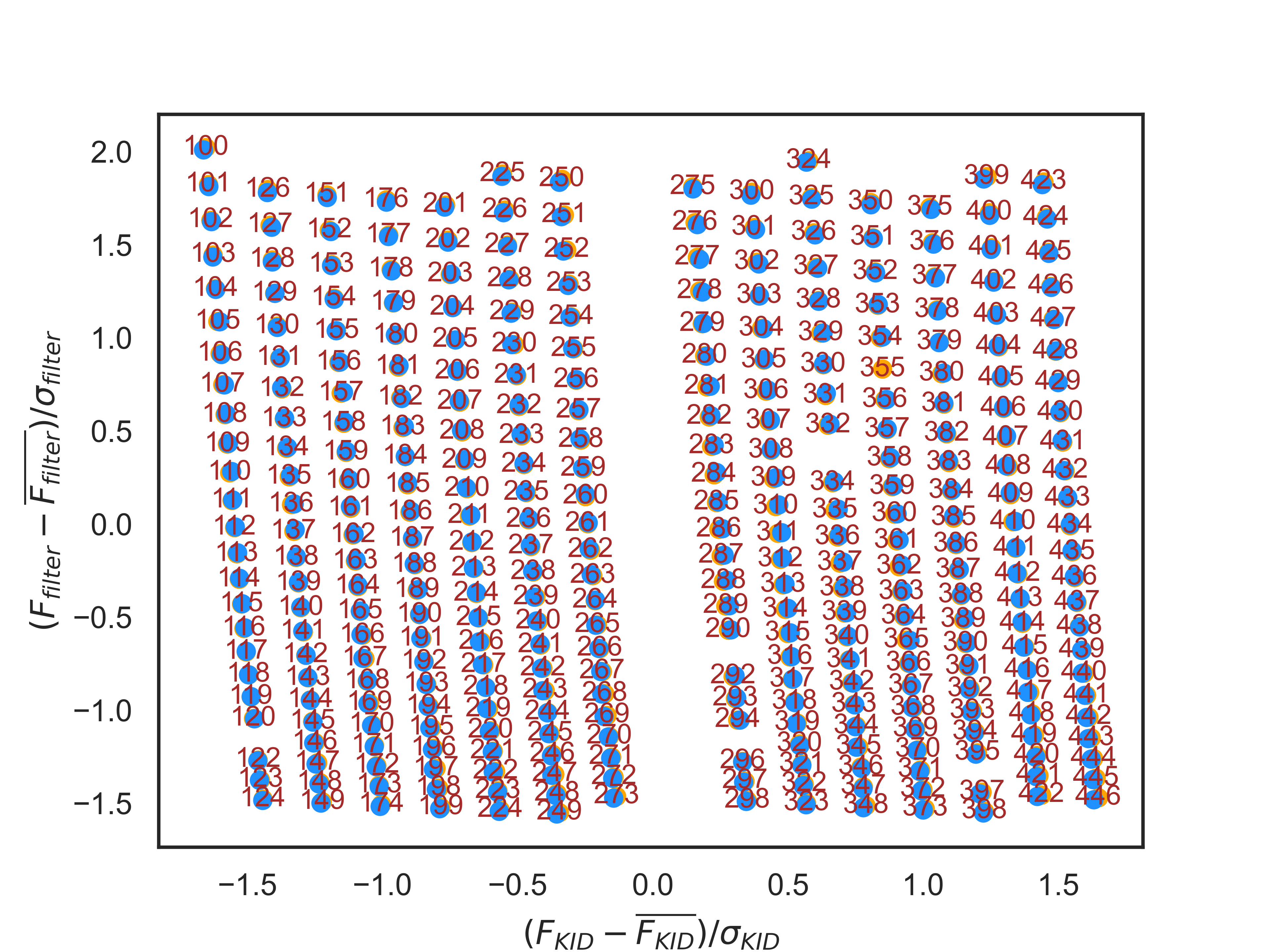}
    \caption{The top panel shows the measured and design values of the filter-KID pairs in blue and orange, respectively. The bottom panel shows the measured pairs after the centering and scaling procedure. The orange numbers are the unique identification numbers for the different filters.}
    \label{fig:point-matching}
\end{figure}

To compare the measured frequencies with the design frequencies of the filters, we make use of the overall statistics of all the filter responses. Each spectral channel is identified by two specific frequencies: the resonant frequency of the filter itself, and the resonant frequency of the coupled kinetic inductance detector (KID). During the design process, special care was taken in choosing these specific frequencies; filters close in frequency space are coupled to more widely spaced KIDs.
Plotting these values in a scatter plot results in an oblique lattice structure where each point is associated with a specific filter-KID pair, as also used in the DESHIMA 2.0 characterization \cite{karatsuDESHIMA202004002026}.
The KID frequencies are measured using a frequency sweep across the readout line. The filter responses are then measured and associated with a specific KID using the continuous-wave mm-submm source. By dividing the filter KID response by that of the wideband KID in front of the filterbank, we compensate for variations in the signal power. We then use a Lorentzian fit to determine the precise resonance frequency.\cite{vanrantwijkMultiplexedReadout1000Pixel2016a}
The resulting measured filter-KID pairs are then compared to the lattice structure of the design, as shown in Figure~\ref{fig:point-matching}. The discrepancy between the two can to a large extent be attributed to fabrication-induced systematic shifts (translation) and scaling in both the filter and KID frequencies.
To compensate for these, we center both lattices around the origin by subtracting the mean frequencies ($F-\overline{F}$) and scaling with the standard deviations ($\sigma$). The resulting lattice overlaps well, and we can match each point in the measured set to one in the design set by identifying the nearest point. This association is then used to compare the measured and designed frequencies. \cite{karatsuDESHIMA202004002026}

\section{Frequency scatter statistics Device 3}
\label{app: Device 3}
\begin{figure*}
    \centering
    \includegraphics[width = 0.8\linewidth]{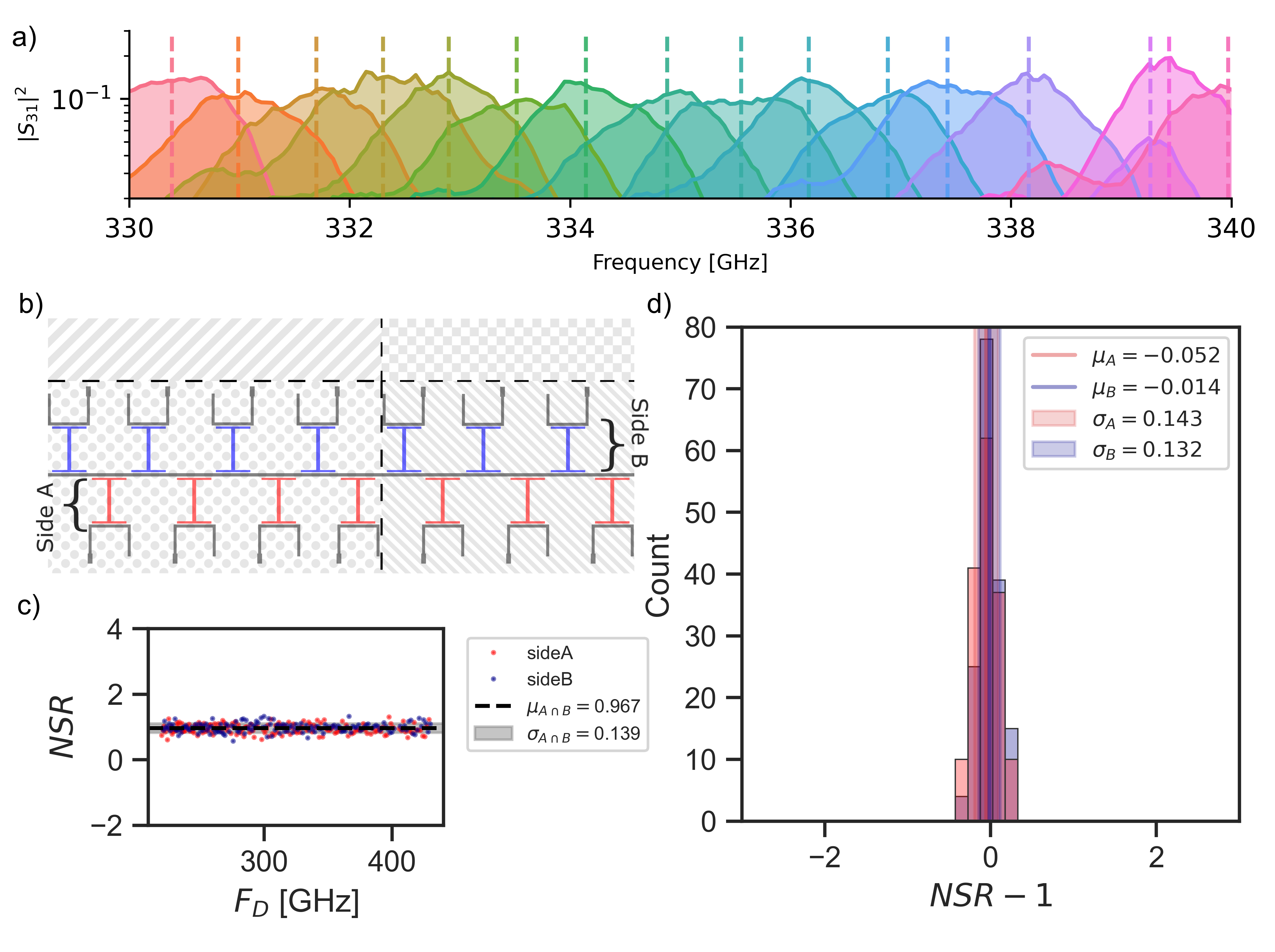}
    \caption{Frequency scatter statistics of D3, a fabrication replicate of D2 used to verify reproducibility. a)~Filter responses in the frequency range 330--340\,GHz. The dashed lines represent the fitted peak frequency of each filter ($F^n_M$). b)~A schematic of the writing pattern. The hatched regions indicate different writing mainfields; the black dashed line marks the mainfield boundaries. As in D2, all filters are fully contained within single mainfields. c)~The normalized spacing ratio (NSR) for each inter-filter gap, where $\mathrm{NSR} = 1$ indicates a perfect match to the design. Each point represents the gap between two adjacent filters. The horizontal dashed line shows the mean NSR ($\mu$) of the filterbank and the shaded band its standard deviation ($\sigma$). d)~A histogram of NSR$\,-\,1$, centering the distributions around zero for the case of perfect frequency spacing. Vertical lines show the mean of each side and the shaded areas their standard deviations. In all panels, colors indicate the two sides of the filterbank (side~A and side~B).\\}
    \label{fig:F-Scatter_LT288}
\end{figure*}

\section*{}
\newpage

\bibliography{Ebeam_fscatter_v5}

\end{document}